\documentclass[
 reprint,
 amsmath,amssymb,
 aps,
]{revtex4-2}

\usepackage{graphicx}
\usepackage{dcolumn}
\usepackage{bm}
\usepackage{lineno}
\usepackage{ulem}
\usepackage{color}
\usepackage[percecent]{overpic}
\usepackage{orcidlink}
\usepackage[nameinlink]{cleveref}
\Crefname{figure}{Fig.}{Figs.}

\begin{document}
\hyphenpenalty=10000
\tolerance=5000
\preprint{APS/123-QED}

\title{
Search for $C\!P$ violation in $D^{+}_{(s)}\rightarrow K^{+}K^{0}_{S}h^{+}h^{-}$~$(h=K,\pi)$ decays\\
and observation of the Cabibbo-suppressed decay $D^{+}_{s}\rightarrow K^{+}K^{-}K^{0}_{S}\pi^{+}$\\
}
\noaffiliation
  \author{H.~K.~Moon\,\orcidlink{0000-0001-5213-6477}} 
  \author{E.~Won\,\orcidlink{0000-0002-4245-7442}} 
  \author{I.~Adachi\,\orcidlink{0000-0003-2287-0173}} 
  \author{H.~Aihara\,\orcidlink{0000-0002-1907-5964}} 
  \author{D.~M.~Asner\,\orcidlink{0000-0002-1586-5790}} 
  \author{H.~Atmacan\,\orcidlink{0000-0003-2435-501X}} 
  \author{V.~Aulchenko\,\orcidlink{0000-0002-5394-4406}} 
  \author{T.~Aushev\,\orcidlink{0000-0002-6347-7055}} 
  \author{R.~Ayad\,\orcidlink{0000-0003-3466-9290}} 
  \author{V.~Babu\,\orcidlink{0000-0003-0419-6912}} 
% ...
\collaboration{The Belle Collaboration}

\begin{abstract}
We search for $C\!P$ violation by measuring a $T$-odd asymmetry in the Cabibbo-suppressed $D^{+}\rightarrow K^{+}K^{0}_{S}\pi^{+}\pi^{-} $ decay, and in the Cabibbo-favored $D^{+}_{s}\rightarrow K^{+}K^{0}_{S}\pi^{+}\pi^{-}$ and $D^{+}\rightarrow K^{+}K^{-}K^{0}_{S}\pi^{+}$ decays.
We use 980 ${\rm fb}^{-1}$ of data collected by the Belle detector running at the KEKB asymmetric-energy $e^{+}e^{-}$ collider.
The $C\!P$-violating $T$-odd parameter ${a}^{T\text{-}\rm{odd}}_{C\!P}$ is measured to be
    ${a}^{T\text{-}\rm{odd}}_{C\!P}(D^{+}\rightarrow K^{+}K^{0}_{S}\pi^{+}\pi^{-})=(0.34\pm0.87\pm0.32)\%,$ 
    ${a}^{T\text{-}\rm{odd}}_{C\!P}(D^{+}_{s}\rightarrow K^{+}K^{0}_{S}\pi^{+}\pi^{-})=(-0.46\pm0.63\pm0.38)\%,$ and ${a}^{T\text{-}\rm{odd}}_{C\!P}(D^{+}\rightarrow K^{+}K^{-}K^{0}_{S}\pi^{+})=
    (-3.34\pm2.66\pm0.35)\%,$ where the first uncertainty is statistical and the second is systematic.
We also report the first observation of the Cabibbo-suppressed decay $D^{+}_{s}\rightarrow K^{+}K^{-}K^{0}_{S}\pi^{+}$.
The branching fraction is measured relative to that of the analogous Cabibbo-favored decay : $\mathcal{B}(D^{+}_{s}\rightarrow K^{+}K^{-}K^{0}_{S}\pi^{+}) / \mathcal{B}(D^{+}_{s}\rightarrow K^{+}K^{0}_{S}\pi^{+}\pi^{-}) = (1.36\pm 0.15\pm 0.04)\%$.
\end{abstract}

\maketitle

In the Standard Model (SM), violation of charge-conjugation and parity ($C\!P$) symmetry in weak decays is described by a quark-mixing phase in the Cabbibo-Kobayashi-Maskawa matrix~\cite{KM1, KM2}.
However, within this theoretical framework, the magnitude of $C\!P$ violation observed so far~\cite{obskaon, obsbelle1, obsbelle2, obsbabar, obslhcb, lhcb1} is too small to explain the large baryon asymmetry in the present universe~\cite{asym}.
Thus, it is important to search for new sources of $C\!P$ violation.

$C\!P$ violation in charm decays is predicted to be especially small: $~10^{-3}$ or less~\cite{charm, kagan}.
Consequently, observing $C\!P$ violation in charm decays could indicate new physics beyond the SM.
The LHCb experiment observed direct $C\!P$ violation in Cabibbo-suppressed (CS) $D^{0}$ decays at the level of 0.2\%~\cite{lhcb1}.
While it is possible to account for this within the SM~\cite{dcp1,dcp3}, new sources of $C\!P$ violation cannot be excluded.

Here we search for $C\!P$ violation in two Cabibbo-favored (\rm{CF}) decays, $D^{+}\rightarrow K^{+}K^{-}K^{0}_{S}\pi^{+}$~\cite{cc} and $D^{+}_{s}\rightarrow K^{+}K^{0}_{S}\pi^{+}\pi^{-}$ [hereafter referred to as $D^{+}$(\rm{CF}) and $D^{+}_{s}$(\rm{CF}), respectively], and a CS decay, $D^{+}\rightarrow K^{+}K^{0}_{S}\pi^{+}\pi^{-}$ [hereafter referred to as $D^{+}$(\rm{CS})].
We also report the first observation of the CS decay $D_{s}^{+}\rightarrow K^{+}K^{-}K^{0}_{S}\pi^{+}$ [hereafter referred to as $D^{+}_{s}$(\rm{CS})], and we measure its branching fraction relative to its CF counterpart $D^{+}_{s}\rightarrow K^{+}K^{0}_{S}\pi^{+}\pi^{-}$.

We search for $T$ violation and $C\!P$ violation by measuring a $T$-odd triple-product asymmetry as follows.
\textcolor{black}{The method was introduced in Refs.~\cite{bigi, london, rosner, durieux} and has been applied to several four-body decays~\cite{tvp1, tvp2, tvp3, tvp4, babar, lhcbnat, aman}.}
We construct the triple product
\textcolor{black}{
\begin{equation}
    C_{T}\equiv \Vec{p}_{K^+} \cdot (\Vec{p}_{\pi^+}\times\Vec{p}_{h^-}),
\end{equation}
where $\vec{p}_{K^+}$, $\vec{p}_{\pi^+}$, and $\vec{p}_{h^-}$ are the momenta of three of the four daughters of a $D_{(s)}^{+}$ decay as measured in the rest frame of the $D_{(s)}^{+}$.}
This quantity changes sign under time reversal, i.e., it is $T$-odd.
We subsequently define $T$-odd asymmetries, $A_T$ and its $C\!P$-conjugate $\bar{A}_T$, as
\begin{equation}
A_{T}=\frac{\Gamma(C_{T}>0)-\Gamma(C_{T}<0)}{\Gamma(C_{T}>0)+\Gamma(C_{T}<0)},
\end{equation}
and
\begin{equation}
\bar{A}_{T}=\frac{\Gamma(-\overline{C}_{T}>0)-\Gamma(-\overline{C}_{T}<0)}{\Gamma(-\overline{C}_{T}>0)+\Gamma(-\overline{C}_{T}<0)}.
\end{equation}
\noindent Here, $\overline{C}_T$ denotes the triple product for the charge-conjugate decay, and the minus sign is included to account for the parity transformation.
The observables $A_T$ and $\bar{A}_T$ can be nonzero due to either $T$ violation or strong phases.
The former would have opposite sign for $A_T$ and $\bar{A}_T$, while the latter would have the same sign~\cite{strong}.
Thus the difference
\begin{equation}
    {a}^{T\text{-}\rm{odd}}_{C\!P}=\frac{1}{2}(A_{T}-\bar{A}_{T})
\end{equation}
removes any effect from strong phases, and a non-zero value would indicate $T$ violation.
As $A_T$ and $\bar{A}_T$ are $C\!P$-conjugate quantities, ${a}^{T\text{-}\rm{odd}}_{C\!P}$ is manifestly $C\!P$-violating.

Previous measurements of the $D^{+}_{(s)}$ decays studied here were made by the {\it BABAR\/\/} experiment, which obtained  ${a}^{T\text{-}\rm{odd}}_{C\!P}[D^{+}(\rm{CS})]=(1.20 \pm 1.00 \pm 0.46)\%$ and ${a}^{T\text{-}\rm{odd}}_{C\!P}[D_{s}^{+}(\rm{CF})]=(-1.36 \pm 0.77 \pm0.34)\%$~\cite{babar}.
We report measurements using a data sample almost twice as large as that used by {\it BABAR\/}.

Our analysis uses data recorded by the Belle experiment, which ran at the KEKB $e^{+}e^{-}$ asymmetric-energy collider \cite{kekb1, kekb2}.
The Belle detector~\cite{belle} is a large-solid-angle magnetic spectrometer that consists of a silicon vertex detector (SVD), a 50-layer central drift chamber (CDC), an array of aerogel threshold Cherenkov counters (ACC), a barrel-like arrangement of time-of-flight scintillation counters (TOF), and an electromagnetic calorimeter consisting of CsI(Tl) crystals.
All these detector components are located inside a superconducting solenoid coil that provides a 1.5 T magnetic field.
An iron flux-return located outside of the coil is instrumented with resistive plate chambers to detect $K^0_L$ mesons and to identify muons.
More information about the detector is provided in Ref.~\cite{belle}.

The data were collected at or near the $\Upsilon(nS)$ $(n=1$--$5)$ resonances and correspond to a total integrated luminosity of 980 ${\rm fb}^{-1}$.
The majority of the data ($711~\rm{fb}^{-1}$) were collected at the $\Upsilon(4S)$ resonance.
We use Monte Carlo (MC) simulation to study sources of background, optimize selection criteria, and calculate event selection efficiencies.
We use {\sc evtgen}~\cite{evtgen} to generate both signal decays and background processes, and {\sc geant3}~\cite{geant3} to model the detector response.
Simulated $D_{(s)}^+\rightarrow K^{+}K^{0}_{S}h^{+}h^{-}$ decays are required to decay uniformly in phase space.
Simulated generic $D$ and $B$ decays include previously measured intermediate resonances.
To avoid bias, all selection criteria and analysis procedures are finalized before examining signal candidates in the data.

To improve track position resolution, all charged tracks are required to have transverse momentum larger than 0.1 GeV/$c$ and at least two associated hits in the SVD in both beam and radial directions.
Charged kaons and pions are identified by the ratio of particle identification (PID) likelihoods $\mathcal{L}_{K}/(\mathcal{L}_{K}+\mathcal{L_{\pi}})$, where $\mathcal{L}_{K}$ and $\mathcal{L_{\pi}}$ are constructed using information from the CDC, TOF, and ACC.
Neutral $K^{0}_{S}$ candidates are reconstructed from the decay chain $K^{0}_{S}\rightarrow \pi^{+}\pi^{-}$.
Identification of the $K^{0}_{S}$ candidates is performed using a neural network algorithm~\cite{nisks} based on kinematic variables of the $K^{0}_{S}$ candidate. 
The invariant mass of the $K^{0}_{S}$ candidates is required to be within $\pm$10 MeV/$c^{2}$ of the nominal value~\cite{PDG}; this corresponds to $\pm5.5\sigma$ in resolution.
For these candidates, a mass-constrained vertex fit is performed.

The $D_{(s)}^{+}$ candidates are reconstructed from three tracks and one $K^{0}_{S}$ candidate.
A vertex fit is performed to the $D_{(s)}^{+}$ candidates.
After this fit, the $D_{(s)}^{+}$ candidates must be consistent with originating from the interaction point (IP).
In particular, we require that the impact parameter with respect to the IP be less than 4.0 cm in the beam direction and less than 2.0 cm in the plane perpendicular to the beam direction.

After this loose selection, two ``peaking" backgrounds that have the same final-state particles as the signal decays are identified.
The first is $D^{+}_{s}\rightarrow K^{+}K^{0}_{S}K^{0}_{S}$, in which both $K^{0}_{S}$ daughters decay to $\pi^{+}\pi^{-}$.
To suppress this background, the invariant mass of the two charged pions originating directly from the $D_{s}^+$ decay is required to be outside the range $\pm 3\sigma$ from the $K^{0}_{S}$ nominal mass~\cite{PDG}.
The second background is $D^{*+}\rightarrow D^{0}\pi^{+}$ followed by $D^{0}\rightarrow K^{+}K^{0}_{S}\pi^{-}$ or $D^{0}\rightarrow K^{+}K^{-}K^{0}_{S}$.
To suppress this background, we require that $\Delta M > 0.15$ GeV/$c^{2}$, where $\Delta M$ is defined as $M(K^{+}K^{0}_{S}\pi^{+}\pi^{-})-M(K^{+}K^{0}_{S}\pi^{-})$ or $M(K^{+}K^{-}K^{0}_{S}\pi^{+})-M(K^{+}K^{-}K^{0}_{S})$.
Here, $M$ denotes the invariant mass of the listed particles.

To veto large combinatorial backgrounds, three quantities are used: the scaled momentum $x_{p}$, the sum of kinematic vertex fit qualities $\Sigma(\chi^{2}/$ndf), and the significance of the $D$ meson decay length $L_{D}/\sigma_{L}$.
The scaled momentum is defined as $x_{p}=p^{*}c/\sqrt{0.25\cdot E^{2}_{\rm{CM}}-{M}^{2}{c}^{4}}$, where $p^{*}$, $E_{\rm{CM}}$, and $M$ are the momentum of the $D_{(s)}^{+}$ candidate in the center-of-mass frame, the total $e^{+}e^{-}$ collision energy in the center-of-mass frame, and the reconstructed invariant mass of the $D_{(s)}^{+}$ candidate, respectively.
The sum $\Sigma(\chi^{2}/$ndf) utilizes the goodness-of-fit $\chi^2$ statistic resulting from the $D^{+}_{(s)}$ production vertex fit and decay vertex fit.
The significance of the $D^{+}_{(s)}$ decay length $L_{D}/\sigma_{L}$ is defined as
\begin{linenomath*}
\begin{align}
    \Vec{L}&=\Vec{r}_{\rm{dec}}-\Vec{r}_{\rm{prod}},\\
    L_{D}&=\Vec{L}\cdot \frac{\Vec{p}}{|\Vec{p}|},\\
    \sigma^{2}_{L}&=\frac{{\Vec{L}}^{T}\cdot(V_{\rm{dec}}+V_{\rm{prod}})\cdot\Vec{L}}{|\Vec{L}|^2},
\end{align}
\end{linenomath*}
where $\Vec{p}$ is the momentum vector of the $D^{+}_{(s)}$, and $\Vec{r}_{\rm{prod}}$ and $\Vec{r}_{\rm{dec}}$ are the position vectors for the production and decay vertices, respectively, each with its corresponding error matrices $V_{\rm{prod}}$ and $V_{\rm{dec}}$.
Signal events typically have larger values of $x_{p}$ and $L_{D}/\sigma_{L}$, and smaller values of $\Sigma(\chi^{2}/$ndf), as compared to background events.

We optimize selection criteria by maximizing the signal significance $S/\sqrt{S+B}$, where $S$ and $B$ are the numbers of signal and background events, respectively, expected in the signal region.
We use MC for signal events and a data sideband for background events.
For $S$, we scale the number of signal events from MC using the known branching fraction~\cite{PDG}.
\textcolor{black}{We optimize selection criteria for the three channels independently, resulting in slightly different selection criteria. 
The optimal values fall within the following ranges: $\Sigma(\chi^{2}/$ndf) $<$ (5 -- 9), $L_{D}/\sigma_{L} >$ (1.4 -- 5.1), and $x_{p} >$ (0.3 -- 0.55).
We account for correlations among these three criteria by optimizing the criteria simultaneously.}

\begin{figure}[t]
\includegraphics[width=1.\linewidth]{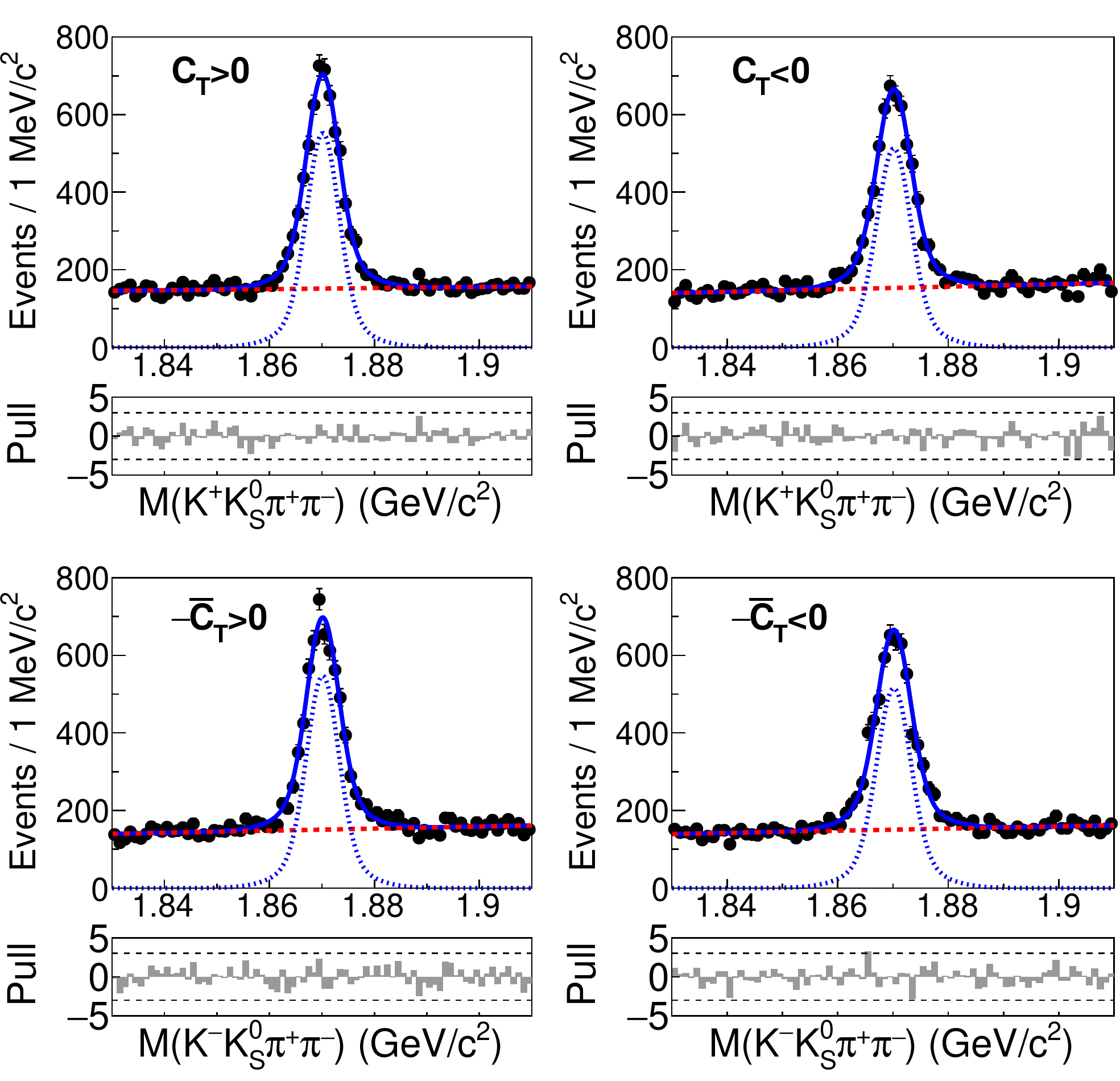}
\caption{\label{fig:simfit1} Fit results for $D^{+}\rightarrow K^{+}K^{0}_{S}\pi^{+}\pi^{-}$ candidates. Dots with error bars show the data; red dashed lines show the background component; blue dotted curves show the signal component; and solid curves show the overall fit result. Pulls are plotted below each mass distribution, with the $\pm3$ level denoted by horizontal lines.} 
\end{figure}

\begin{figure}[hbtp]
\includegraphics[width=1.\linewidth]{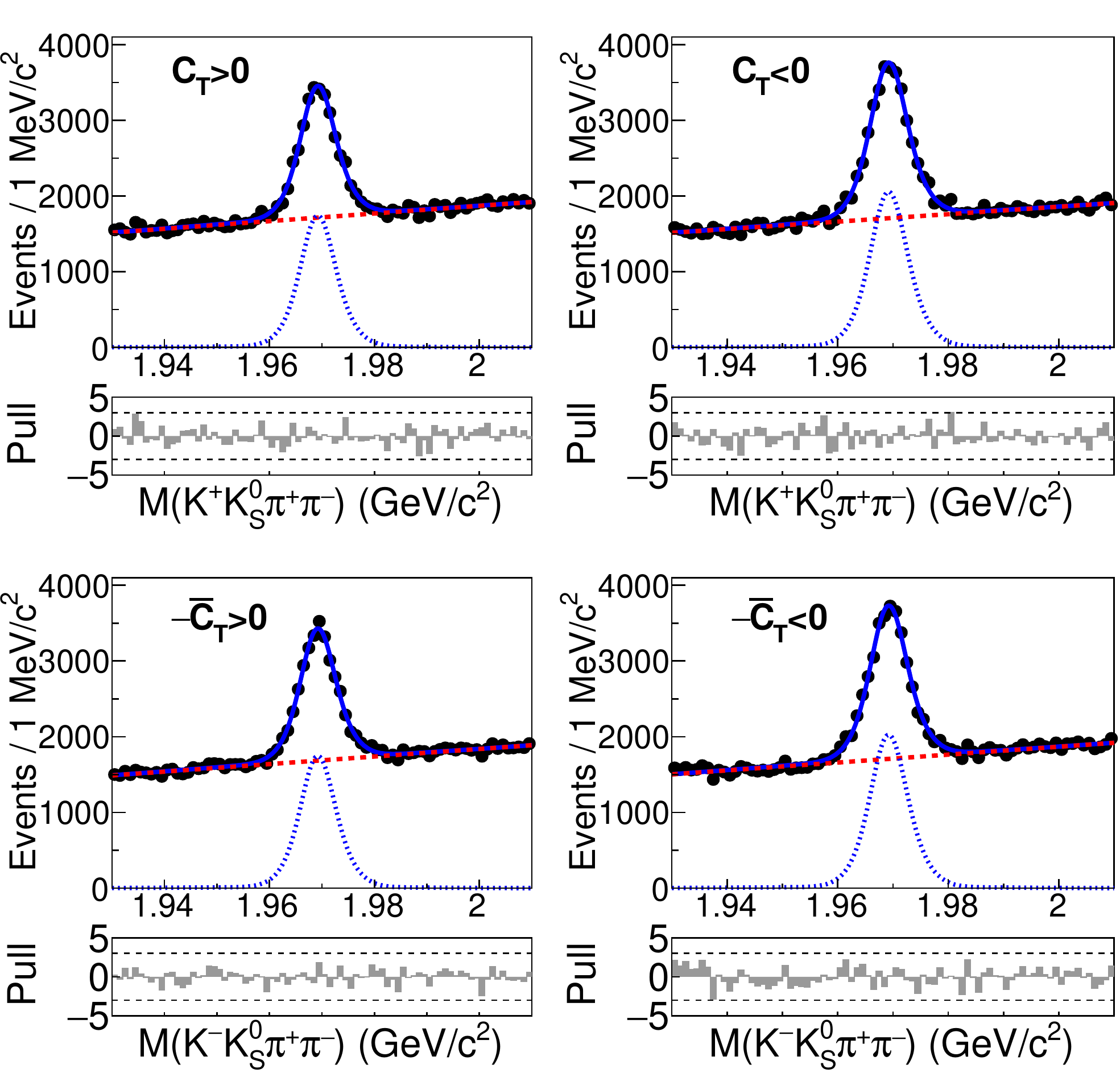}
\caption{\label{fig:simfit2} Fit results for $D_{s}^{+}\rightarrow K^{+}K^{0}_{S}\pi^{+}\pi^{-}$ candidates. Dots with error bars show the data; red dashed lines show the background component; blue dotted curves show the signal component; and solid curves show the overall fit result. Pulls are plotted below each mass distribution, with the $\pm3$ level denoted by horizontal lines.}
\end{figure}

\begin{figure}[ht]
\includegraphics[width=1.\linewidth]{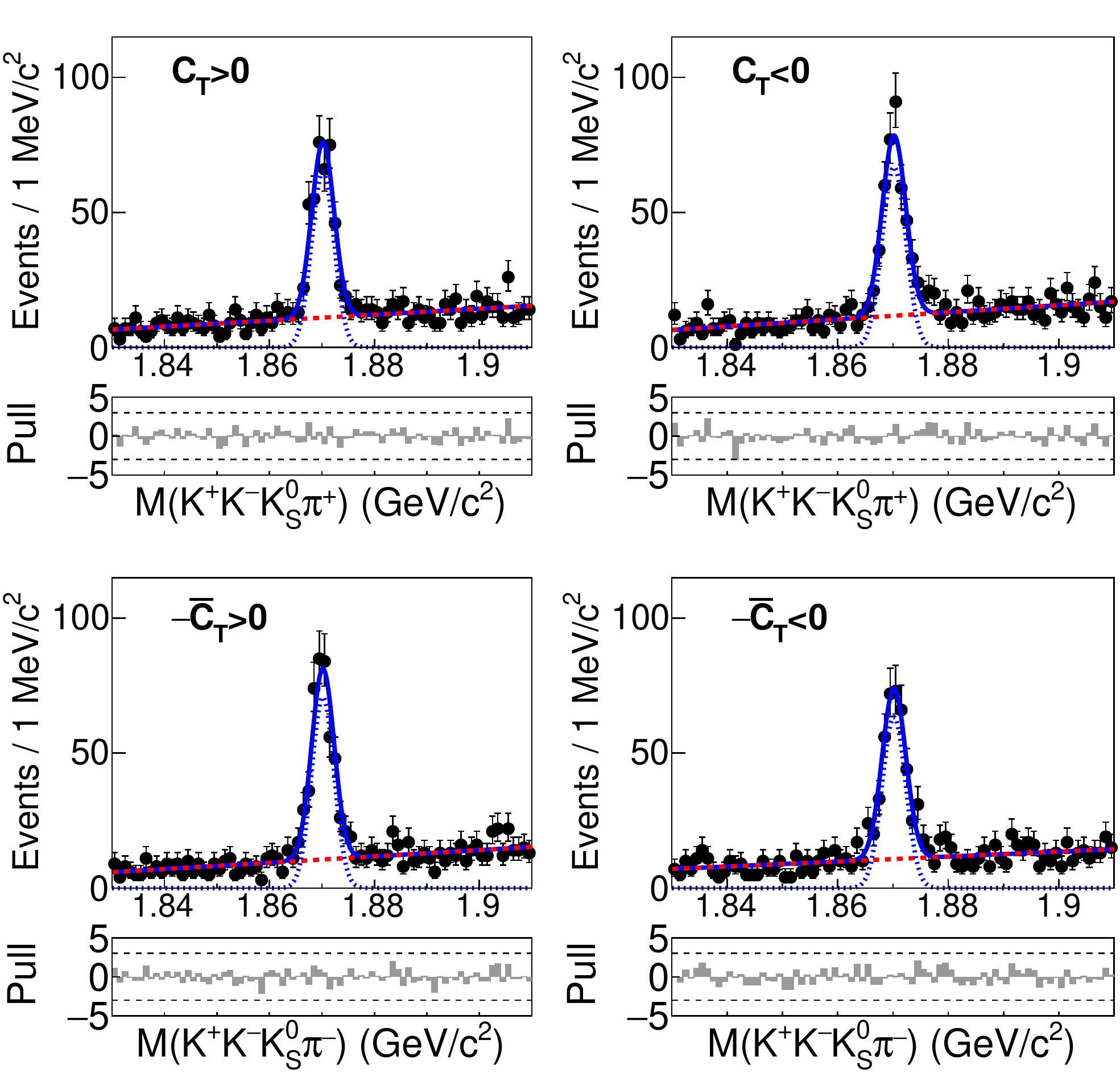}
\caption{\label{fig:simfit3} Fit results for $D^{+}\rightarrow K^{+}K^{-}K^{0}_{S}\pi^{+}$ candidates. Dots with error bars show the data; red dashed lines show the background component; blue dotted curves show the signal component; and solid curves show the overall fit result. Pulls are plotted below each mass distribution, with the $\pm3$ level denoted by horizontal lines.}
\end{figure}

The invariant mass distributions of signal candidates after applying all selection criteria are shown in Figs.~\ref{fig:simfit1}--\ref{fig:simfit3}.
For each channel, events are divided into four subsamples, depending on the $D$ charge and sign of $C_T$ and $\overline{C}_{T}$ values.
The four signal yields are related to the $T$-odd observable $A_T$ and $C\!P$-violating parameter ${a}^{T\text{-}\rm{odd}}_{C\!P}$ as follows:
\begin{linenomath*}
\begin{align}
N(C_{T}>0)&=\frac{N(D_{(s)}^{+})}{2}(1+A_{T}),\\
N(C_{T}<0)&=\frac{N(D_{(s)}^{+})}{2}(1-A_{T}),\\
N(-\overline{C}_{T}>0)&=\frac{N(D_{(s)}^{-})}{2}(1+A_{T}-2{a}^{T\text{-}\rm{odd}}_{C\!P}),\\
N(-\overline{C}_{T}<0)&=\frac{N(D_{(s)}^{-})}{2}(1-A_{T}+2{a}^{T\text{-}\rm{odd}}_{C\!P}).    
\end{align}
\end{linenomath*}

We determine $N(D_{(s)}^{+})$, $N(D_{(s)}^{-})$, $A_T$, and ${a}^{T\text{-}\rm{odd}}_{C\!P}$ by performing a binned maximum likelihood fit, simultaneously to the invariant mass distributions of the four subsamples.
The signal component is described by the superposition of two Gaussian functions with a common mean value.
The background component is modeled with a straight line.
We use a common signal probability density function (PDF) and four independent background PDFs for the subsamples.
All parameters of the PDFs are free to vary.
The asymmetries $A_T$ and ${a}^{T\text{-}\rm{odd}}_{C\!P}$ are directly extracted from the fit.
To validate our method, we extract $A_T$ and ${a}^{T\text{-}\rm{odd}}_{C\!P}$ from six independent MC samples where no $T$-violation is expected.
For all MC samples, the extracted asymmetries are consistent with zero.

Projections of the fit result are superimposed on the data in Figs.~\ref{fig:simfit1}--\ref{fig:simfit3}.
The normalized residuals (``pulls") are plotted below the distributions and are calculated as $(N_{\rm{data}}-N_{\rm{fit}})/\sigma_{N_{\rm{data}}}$.
Here $N_{\rm{data}}$, $N_{\rm{fit}}$, and $\sigma_{N_{\rm{data}}}$ are the yield, yield predicted by the fitted PDF, and the error on the yield, respectively.
The fitted results for $A_T$ and ${a}^{T\text{-}\rm{odd}}_{C\!P}$ are listed in Table~\ref{tab:acp_1}.

\begin{table}[t]
\caption{Results of $A_{T}$ and ${a}^{T\text{-}\rm{odd}}_{C\!P}$ measurements. The uncertainties listed are statistical.\label{tab:acp_1} }
\begin{ruledtabular}
\begin{tabular}{ccc}
Mode & ${A}_{T}$ (\%) & ${a}^{T\text{-}\rm{odd}}_{C\!P}$  (\%)\\
\hline
$D^{+}\rightarrow K^{+}K^{0}_{S}\pi^{+}\pi^{-}$& \hspace{0.25cm}$(3.67 \pm 1.23)$ & \hspace{0.25cm}$(0.34 \pm 0.87)$ \\
$D_{s}^{+}\rightarrow K^{+}K^{0}_{S}\pi^{+}\pi^{-}$&$(-8.31 \pm 8.89)$ &$(-0.46 \pm 0.63)$ \\
$D^{+}\rightarrow K^{+}K^{-}K^{0}_{S}\pi^{+}$& $(-1.40 \pm 4.23)$ &$(-3.34 \pm 2.66)$ \\
\end{tabular}
\end{ruledtabular}
\end{table}


The main sources of systematic uncertainty are listed in Table~\ref{tab:sys1} and evaluated as follows.
Possible bias resulting from the choice of signal shape is checked by fitting for ${a}^{T\text{-}\rm{odd}}_{C\!P}$ using alternative shapes.
For these shapes we try a Gaussian function, the superposition of a Gaussian function and an asymmetric Gaussian function, and the superposition of two asymmetric Gaussian functions.
The largest deviation of ${a}^{T\text{-}\rm{odd}}_{C\!P}$ from the nominal result is taken as the systematic uncertainty.

The systematic uncertainty from possible detector bias is checked by measuring ${a}^{T\text{-}\rm{odd}}_{C\!P}$ for control sample $D^{+}\rightarrow K^{0}_{S}\pi^{+}\pi^{+}\pi^{-}$.
This control sample is CF decay and is expected to have ${a}^{T\text{-}\rm{odd}}_{C\!P}$ value consistent with zero with small statistical uncertainty.
We obtain ${a}^{T\text{-}\rm{odd}}_{C\!P}=(-0.32\pm0.27$)\% for $D^{+}\rightarrow K^{0}_{S}\pi^{+}\pi^{+}\pi^{-}$.
We assign this central values of ${a}^{T\text{-}\rm{odd}}_{C\!P}$ as the systematic uncertainty due to possible detector bias.

\textcolor{black}{We also check for possible biases due to $C_{T}$ resolution and differences in reconstruction efficiency among the four subsamples of each mode.}
These uncertainties are evaluated by taking the difference between generated and reconstructed values of ${a}^{T\text{-}\rm{odd}}_{C\!P}$ for the signal MC samples.
The total systematic uncertainties are evaluated as the sum in quadrature of all individual contributions and are also listed in Table~\ref{tab:sys1}. The results for ${a}^{T\text{-}\rm{odd}}_{C\!P}$ are listed in Table~\ref{tab:acp} along with the corresponding signal yields.

\begin{table}[phtb]
\caption{\label{tab:sys1} Contributions to the absolute systematic uncertainty for ${a}^{T\text{-}\rm{odd}}_{C\!P}$ in units of \% for each mode.}
\begin{ruledtabular}
\begin{tabular}{lccc}
Sources & $D^{+}$(\rm{CS}) & $D^{+}_{s}$(\rm{CF}) & $D^{+}$(\rm{CF})\\
\hline
Fit model & 0.01 & 0.02 & 0.12\\
Detector bias & 0.32  & 0.32 & 0.32 \\
$C_T$, $\overline{C}_{T}$ efficiency and resolution & 0.03& 0.20& 0.06\\
\hline
Total & 0.32 & 0.38 & 0.35 \\
\end{tabular}
\end{ruledtabular}
\end{table}

\begin{table}[phtb]
\caption{Fitted signal yields and ${a}^{T\text{-}\rm{odd}}_{C\!P}$ values. The first uncertainties are statistical and the second are systematic.\label{tab:acp} }
\begin{ruledtabular}
\begin{tabular}{ccc}
Mode & $N(D^{+}_{(s)})$ & ${a}^{T\text{-}\rm{odd}}_{C\!P}$ (\%)\\
\hline
$D^{+}\rightarrow K^{+}K^{0}_{S}\pi^{+}\pi^{-}$& $18632\pm214$ & \hspace{0.25cm}$(0.34 \pm 0.87 \pm 0.32)$ \\
$D_{s}^{+}\rightarrow K^{+}K^{0}_{S}\pi^{+}\pi^{-}$& $70080\pm676$ &$(-0.46 \pm 0.63 \pm 0.38)$ \\
$D^{+}\rightarrow K^{+}K^{-}K^{0}_{S}\pi^{+}$&$1425\pm44$ &$(-3.34 \pm 2.66 \pm 0.35)$
\end{tabular}
\end{ruledtabular}
\end{table}

We also perform a search for the CS decay $D_{s}^{+}\rightarrow K^{+}K^{-}K^{0}_{S}\pi^{+}$.
We suppress peaking background from $D^{*+}\rightarrow D^{0}\pi^{+}$, $D^{0}\rightarrow K^{+}K^{-}K^{0}_{S}$ by requiring $\Delta M > 0.15$ GeV/$c^{2}$.
As done for the ${a}^{T\text{-}\rm{odd}}_{C\!P}$ measurement, we suppress backgrounds using the variables $x_{p}$, $\Sigma(\chi^{2}/$ndf), and the significance of the $D$ meson decay length, $L_{D}/\sigma_{L}$.
The selection criteria are chosen to maximize the ratio $S/\sqrt{B}$, where $S$ and $B$ are the expected yields of signal and background events in the signal region based on MC simulation.
To normalize the sensitivity of our search, we use the fitted yield of CF $D^{+}_{s}\rightarrow K^{+}K^{0}_{S}\pi^{+}\pi^{-}$ decays; dividing the yield of $D^{+}_{s}\rightarrow K^{+}K^{-}K^{0}_{S}\pi^{+}$ by this yield and the ratio of efficiencies gives the ratio of branching fractions.

\begin{figure}[htbp]
\includegraphics[width=1.0\linewidth]{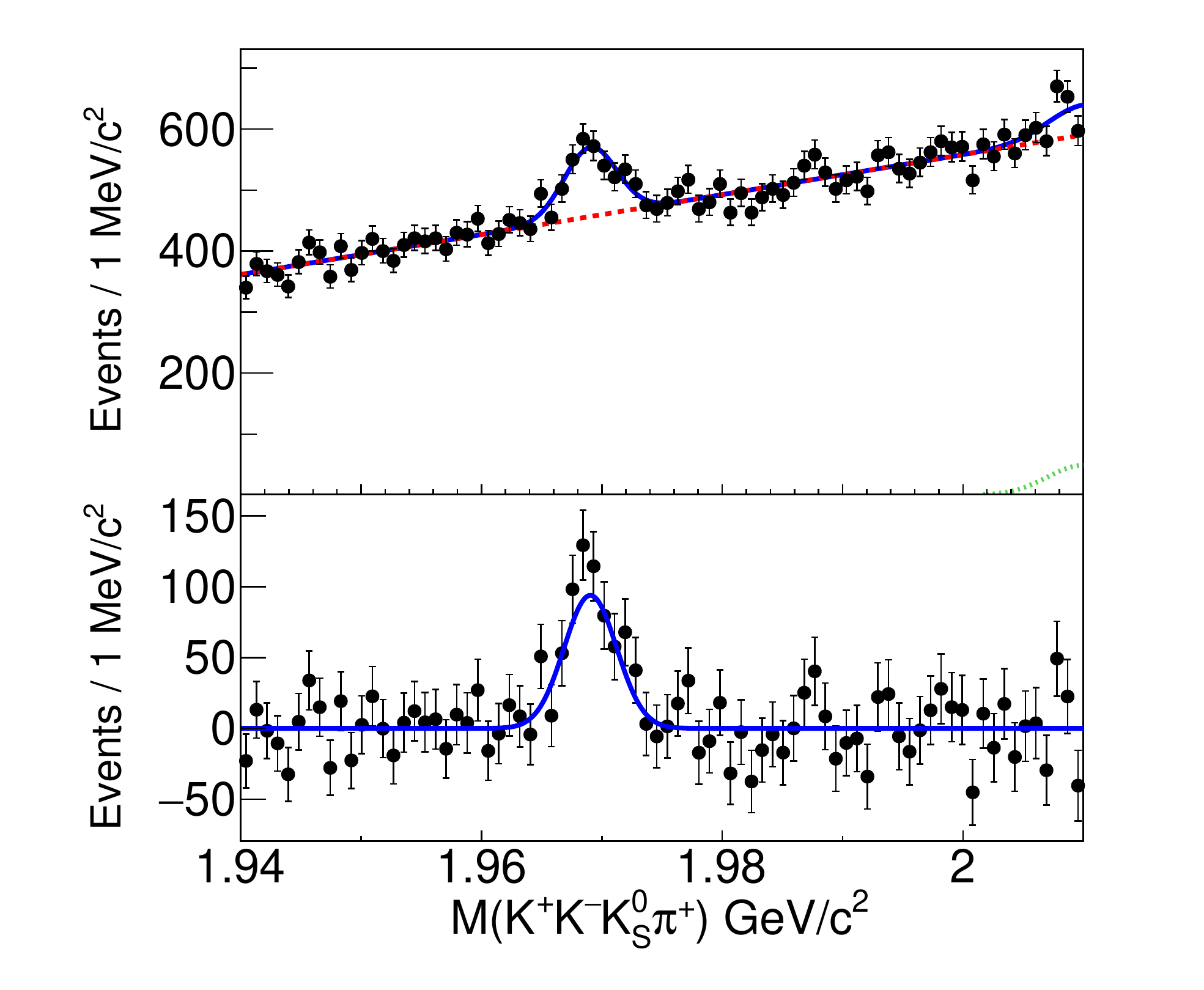}
\caption{\label{fig:obs} The $M(K^{+}K^{-}K^{0}_{S}\pi^{+})$ distribution for signal candidates with the fit result superimposed. Circles with error bars show the data, and the solid contour shows the overall fit result. In the top plot, the red dashed contour shows the combinatorial background, and the green dotted contour shows the $D^{*+}$ background. In the lower plot, these background components have been subtracted.
The background component is subtracted in the lower histogram.}
\end{figure}

The distribution of $M(K^+ K^- K_S^0 \pi^+)$ after applying all selection criteria is shown in Fig.~\ref{fig:obs}.
A clear peak at the mass of the $D_s^+$~\cite{PDG} is observed.
To obtain the signal yield, we perform a maximum likelihood fit to the $M(K^+ K^- K_S^0 \pi^+)$ distribution.
A Gaussian function and a straight line are used to describe the shapes of the signal and combinatorial background, respectively.
The shape of residual $D^{*+}$ background is taken to be a Gaussian, with the mean and width fixed to MC values.
The signal yield obtained is $645\pm70$.
The statistical significance of the $D_s^+$(CS) signal is 9.3$\sigma$, calculated using the difference in the log likelihoods $\sqrt{-2\rm{ln}(\mathcal{L_{\rm{0}}/L_{\rm{max}}})}$.
Here, $\mathcal{L_{\rm{max}}}$ and $\mathcal{L}_{\rm{0}}$ are the likelihood values of the fit to the $M(K^{+}K^{-}K^{0}_{S}\pi^{+})$ spectrum with and without including the signal PDF, respectively.
In order to estimate the signal significance including the additive systematic uncertainties, fits using alternative PDFs for signal and background are performed as discussed below.
The minimum value of signal significance we obtain is 9.2$\sigma$.
To be conservative, we use this value as the signal significance with systematic uncertainties included.

To take into account variation in reconstruction efficiencies due to unknown intermediate resonances, we correct the fitted signal yield for efficiency in bins of five-dimensional (5D) phase space.
We only use events in a signal region defined as $|M(K^{+}K^{0}_{S}h^{-}\pi^{+})-m(D_s^{+})|<$ 10~MeV/$c^{2}$.
These bins consist of the invariant masses of pairs of final-state particles.
Such a method has been used in other analyses of four-body decays of $D$ mesons~\cite{effcor2}.
The 5D phase space is divided into 243 bins (i.e., $3\times3\times3\times3\times3$), which is well-matched to the structure of the efficiency distribution obtained from the signal MC sample.
\textcolor{black}{The binning is first chosen to minimize the efficiency variations within the bins.
Subsequently, we adjust the binning to ensure there are no empty bins. }
We calculate the corrected signal yield $N^{\rm{corr}}$ as 
\begin{equation}
\label{eq:corr}
    N^{\rm{corr}} = \sum_{i} \frac{N_{i}^{\rm{tot}}-N^{\rm{bkg}}\cdot f_{i}^{\rm{bkg}}}{\epsilon_{i}}.
\end{equation}
Here, $N_{i}^{\rm{tot}}$ and $\epsilon_{i}$ are the total number of events and the reconstruction efficiency, respectively, for the $i$th bin, and $N^{\rm{bkg}}$ is the overall number of background events for all bins together.
The fraction of background events in bin $i$ ($f_{i}^{\rm{bkg}}$) and $\epsilon_{i}$ are obtained from MC simulation.
The uncertainties on each term in Eq.~(\ref{eq:corr}) are propagated to obtain the overall uncertainty on $N^{\rm{corr}}$.
Only the PID requirement for a single charged track is different between the final state particles of the signal and normalization modes.
To account for a small difference in PID efficiency between data and MC simulation, a correction for PID is included in the efficiency calculation.
The correction factor is obtained from a $D^{*+}\rightarrow D^{0}\pi^{+}$, $D^{0}\rightarrow K^{-}\pi^{+}$ control sample.
To account for the difference in the momentum spectra of the daughter tracks between the analysis mode and the control sample, the daughter tracks are divided into 384 bins according to the momentum and polar angle (32 momentum bins and 12 polar angle bins).
We obtain the efficiency-corrected signal yields as listed in Table~\ref{tab:br}, and the relative branching fraction $\mathcal{B}[D^{+}_{s}$(CS)]/$\mathcal{B}[D^{+}_{s}$(CF)] = $(1.36\pm 0.15)$\%, where the uncertainty is statistical only.

\begin{table}[htbp]
\caption{\label{tab:br}%
Fitted signal yields ($N^{\rm{sig}}$) and efficiency-corrected signal yields ($N^{\rm{corr}}$) for $D^{+}_{s}\rightarrow K^{+}K^{0}_{S}\pi^{+}\pi^{-}$ and $D^{+}_{s}\rightarrow K^{+}K^{-}K^{0}_{S}\pi^{+}$.}
\begin{ruledtabular}
\begin{tabular}{ccc}
\textrm{Decay mode}&
\textrm{$N^{\rm{sig}}$}&
\textrm{$N^{\rm{corr}}(\times10^2)$}\\
\colrule
$D^{+}_{s}\rightarrow K^{+}K^{0}_{S}\pi^{+}\pi^{-}$& $70080\pm 676$ & $10782\pm 104$\\
$D^{+}_{s}\rightarrow K^{+}K^{-}K^{0}_{S}\pi^{+}$& $645\pm 70$ & $146\pm 15$\\
\end{tabular}
\end{ruledtabular}
\end{table}

\begin{table}
\caption{\label{tab:sys2} Contributions to the fractional systematic uncertainty for the ratio of branching fractions $\mathcal{B}({D^{+}_{s}}$(CS))/$\mathcal{B}({D^{+}_{s}}$(CF)) in \%.}
\begin{ruledtabular}
\begin{tabular}{ll}
Sources & (\%)\\
\hline
PID efficiency correction & 1.6\\
Efficiency correction (binning)& 0.7\\
Efficiency correction (intermediate resonances)& 0.5\\
PDF model & 1.8\\
\hline
Total & 2.6\\
\end{tabular}
\end{ruledtabular}
\end{table}

The main sources of systematic uncertainty are listed in Table~\ref{tab:sys2} and evaluated as follows.
Since the correction for the difference in PID efficiencies between data and MC is included in the calculation of the signal yield correction, the uncertainty of the correction is evaluated.
We assign 1.6\% as the systematic uncertainty for this contribution.

The uncertainty from the efficiency correction method is checked by using different binnings of the 5D phase space.
The largest deviation from the nominal value is assigned as the uncertainty.

To check for any remaining bias due to possible intermediate resonances, we generate MC samples of signal decays proceeding through intermediate resonances and re-calculate $N^{\rm corr}$ using Eq.~(\ref{eq:corr}).
The largest deviation observed in the ratio of branching fractions with respect to our nominal values is 0.5\%, and we assign this value as a systematic uncertainty.

Systematic uncertainties associated with the signal PDF, combinatorial background PDF, and $D^{*0}$ background shape in $D^{+}_{s}$(CS) are considered as follows.
(1) We try alternative signal shapes as in the ${a}^{T\text{-}\rm{odd}}_{C\!P}$ measurement.
(2) The combinatorial background PDF is replaced by a second-order polynomial.
Using the combination of the alternative shapes for signal and background PDFs, we recalculate the ratio of branching fractions.
We assign the root-mean-square value of the variation as the uncertainty.
(3) The $D^{*0}$ background shape in $D^{+}_{s}$(CS) is checked by varying the fixed parameters.
We refit the yield of $D^{+}_{s}$(CS) signal yields for 500 different sets of parameters.
The root-mean-square value of the variation is taken as the systematic uncertainty.
The uncertainties obtained from (1), (2), and (3) are summed in quadrature and assigned as the systematic uncertainty due to the PDF modeling.
The total systematic uncertainty is calculated as the sum in quadrature of the uncertainties from all sources, as listed in Table~\ref{tab:sys2}.

In summary, using 980 ${\rm fb}^{-1}$ of data collected with the Belle detector, we measure the $C\!P$- and $T$-violating parameter ${a}^{T\text{-}\rm{odd}}_{C\!P}$ for the decays $D^{+}\rightarrow K^{+}K^{0}_{S}\pi^{+}\pi^{-}$, $D^{+}_{s}\rightarrow K^{+}K^{0}_{S}\pi^{+}\pi^{-}$, and $D^{+}\rightarrow K^{+}K^{-}K^{0}_{S}\pi^{+}$.
The results are
\begin{linenomath*}
\begin{align}
    {a}^{T\text{-}\rm{odd}}_{C\!P}(D^{+}\rightarrow K^{+}K^{0}_{S}\pi^{+}\pi^{-})&=\phantom{-}
    (0.34\pm0.87\pm0.32)\%\nonumber\\
    {a}^{T\text{-}\rm{odd}}_{C\!P}(D^{+}_{s}\rightarrow K^{+}K^{0}_{S}\pi^{+}\pi^{-})&=
    (-0.46\pm0.63\pm0.38)\%\nonumber\\
    {a}^{T\text{-}\rm{odd}}_{C\!P}(D^{+}\rightarrow K^{+}K^{-}K^{0}_{S}\pi^{+})&=
    (-3.34\pm2.66\pm0.35)\%\nonumber,
\end{align}
\end{linenomath*}
where the first and second uncertainties are statistical and systematic, respectively.
The results are the most precise to date and are consistent with no $C\!P$ violation in these modes.

We also report the first observation of the CS decay $D^{+}_{s}\rightarrow K^{+}K^{-}K^{0}_{S}\pi^{+}$ with a signal significance of 9.2$\sigma$.
The branching fraction for $D_{s}^{+}\rightarrow K^{+}K^{-}K^{0}_{S}\pi^{+}$ relative to that for the CF decay $D_{s}^{+}\rightarrow K^{+}K^{0}_{S}\pi^{+}\pi^{-}$ is measured to be
\begin{equation}
    \frac{\mathcal{B}(D^{+}_{s}\rightarrow K^{+}K^{-}K^{0}_{S}\pi^{+})}{\mathcal{B}(D_{s}^{+}\rightarrow K^{+}K^{0}_{S}\pi^{+}\pi^{-})} = (1.36\pm 0.15\pm0.04)\%,\nonumber\\
\end{equation}
where the first and second uncertainties are statistical and systematic, respectively.
Inserting the world average value for the branching fraction of the normalization mode, $\mathcal{B}(D^{+}_{s}\rightarrow K^{+}K^{0}_{S}\pi^{+}\pi^{-})=(0.95\pm0.08)\%$~\cite{PDG}, we obtain
\begin{linenomath*}
\begin{align}
\mathcal{B}(D^{+}_{s}\rightarrow K^{+}K^{-}K^{0}_{S}\pi^{+}) = \hspace{3cm}\nonumber
\\(1.29\pm 0.14\pm 0.04\pm 0.11)\times {10}^{-4},\nonumber
\end{align}
\end{linenomath*}
where the the third uncertainty is due to uncertainty in the branching fraction of the normalization mode.

This work, based on data collected using the Belle detector, which was
operated until June 2010, was supported by 
the Ministry of Education, Culture, Sports, Science, and
Technology (MEXT) of Japan, the Japan Society for the 
Promotion of Science (JSPS), and the Tau-Lepton Physics 
Research Center of Nagoya University; 
the Australian Research Council including grants
DP180102629, 
DP170102389, 
DP170102204, 
DE220100462, 
DP150103061, 
FT130100303; 
Austrian Federal Ministry of Education, Science and Research (FWF) and
FWF Austrian Science Fund No.~P~31361-N36;
the National Natural Science Foundation of China under Contracts
No.~11675166,  
No.~11705209;  
No.~11975076;  
No.~12135005;  
No.~12175041;  
No.~12161141008; 
Key Research Program of Frontier Sciences, Chinese Academy of Sciences (CAS), Grant No.~QYZDJ-SSW-SLH011; 
Project ZR2022JQ02 supported by Shandong Provincial Natural Science Foundation;
the Ministry of Education, Youth and Sports of the Czech
Republic under Contract No.~LTT17020;
the Czech Science Foundation Grant No. 22-18469S;
Horizon 2020 ERC Advanced Grant No.~884719 and ERC Starting Grant No.~947006 ``InterLeptons'' (European Union);
the Carl Zeiss Foundation, the Deutsche Forschungsgemeinschaft, the
Excellence Cluster Universe, and the VolkswagenStiftung;
the Department of Atomic Energy (Project Identification No. RTI 4002) and the Department of Science and Technology of India; 
the Istituto Nazionale di Fisica Nucleare of Italy; 
National Research Foundation (NRF) of Korea Grant
Nos.~2016R1\-D1A1B\-02012900, 2018R1\-A2B\-3003643,
2018R1\-A6A1A\-06024970, RS\-2022\-00197659,
2019R1\-I1A3A\-01058933, 2021R1\-A6A1A\-03043957,
2021R1\-F1A\-1060423, 2021R1\-F1A\-1064008, 2022R1\-A2C\-1003993;
Radiation Science Research Institute, Foreign Large-size Research Facility Application Supporting project, the Global Science Experimental Data Hub Center of the Korea Institute of Science and Technology Information and KREONET/GLORIAD;
the Polish Ministry of Science and Higher Education and 
the National Science Center;
the Ministry of Science and Higher Education of the Russian Federation, Agreement 14.W03.31.0026, 
and the HSE University Basic Research Program, Moscow; 
University of Tabuk research grants
S-1440-0321, S-0256-1438, and S-0280-1439 (Saudi Arabia);
the Slovenian Research Agency Grant Nos. J1-9124 and P1-0135;
Ikerbasque, Basque Foundation for Science, Spain;
the Swiss National Science Foundation; 
the Ministry of Education and the Ministry of Science and Technology of Taiwan;
and the United States Department of Energy and the National Science Foundation.
These acknowledgements are not to be interpreted as an endorsement of any
statement made by any of our institutes, funding agencies, governments, or
their representatives.
We thank the KEKB group for the excellent operation of the
accelerator; the KEK cryogenics group for the efficient
operation of the solenoid; and the KEK computer group and the Pacific Northwest National
Laboratory (PNNL) Environmental Molecular Sciences Laboratory (EMSL)
computing group for strong computing support; and the National
Institute of Informatics, and Science Information NETwork 6 (SINET6) for
valuable network support.
E.~W. and H.~M. are partially supported by NRF grants 2019H1D3A1A01101787 and 2022R1A2B5B02001535.


\begin{thebibliography}{}
\bibitem{cc} Here and throughout the paper, charge-conjugate modes are implied unless stated otherwise.
\end{thebibliography}
{}
\end{document}